\documentclass[twocolumn]{aastex631}

\newcommand{\HI}{{\rm HI}}
\newcommand{\mpc}{{\rm Mpc}}

\usepackage{natbib}
\usepackage{amsmath} 
\defcitealias{chowdhury2024hi}{CH24}
\newcommand{\re}[1]{{{#1}}}
\begin{document}
% \label{firstpage}
% \pagerange{\pageref{firstpage}--\pageref{lastpage}}
% \maketitle

% \title{The Evolving Power Spectrum in parameter studies of the Epoch of Reionization 21-cm signal}
% \title{Parameter inference for the Light-cone Epoch of Reionization 21-cm signal}

\title{Constraining the $z \approx 1$ neutral hydrogen (HI) distribution}

% \title{Accurate parameter inference for the Light-cone Epoch of Reionization 21-cm signal using the SKA-Low}

\correspondingauthor{Minal Chhabra}
\email{minal99chhabra@gmail.com}

\author[0009-0006-2294-3490]{Minal Chhabra}
\affiliation{Department of Physics, Indian Institute of Technology Kharagpur, Kharagpur 721 302, India}

\author[0000-0001-9816-5070]{Raghunath Ghara}
\affiliation{Department of Physics, Indian Institute of Technology Kharagpur, Kharagpur 721 302, India}

\author[0000-0002-2350-3669]{Somnath Bharadwaj}
\affiliation{Department of Physics, Indian Institute of Technology Kharagpur, Kharagpur 721 302, India}

\begin{abstract}
We constrain the $z \approx 1$ HI distribution by jointly modeling two independent, existing observations of the 21-cm signal, the HI density parameter $\Omega_\HI$ measured using uGMRT by stacking the 21-cm emission from blue, star-forming galaxies, and the 21-cm autocorrelation power spectrum (PS) measured using CHIME. We assign HI to the dark matter halos in a cosmological simulation using an HI mass-halo mass (HIHM) relation with three free parameters whose values we estimate by performing a joint Bayesian inference comparing the simulated $\Omega_\HI$ and 21-cm PS with the measurements. We use the inferred HIHM posterior to simulate the HI distribution and predict the $z \approx 1$ HI mass function (HIMF). 
We find that the HIMF remains nearly constant at low HI masses $( \lesssim 3.6 \times 10^9 )$, and it declines rapidly for larger HI masses. Around $\sim 90$ percent of the total HI gas is contained in the mass range $M_\HI \in [2 \times 10^9, \, 4 \times 10^{11}] \, M_\odot$. Our estimates predict a larger abundance of high mass HI galaxies than predicted by earlier observations and hydrodynamical simulations. We expect these results to be useful in understanding galaxy evolution and star formation.
\end{abstract}

\section{Introduction}
\label{sec:intro}
The evolution of the atomic neutral hydrogen (HI) contained in galaxies is of considerable interest, not only to understand star formation \citep{binney2011galactic}, but also for cosmological HI 21-cm Intensity Mapping (IM) \citep*{bharadwaj2001using}. It is observed that the cosmic star formation rate (SFR) undergoes a peak between $z=1-3$ when the majority of the stars are formed and then declines exponentially at $z<1$ \citep{madau2014cosmic}. In a sequence of processes, the gas accreted from the Intergalactic medium (IGM) cools and collapses to form HI, then molecular hydrogen clouds that finally fragment to form stars. In this work, we consider the $z \approx 1$ HI mass function (HIMF), which quantifies the comoving number density of galaxies with a given HI mass. The direct estimates of the HIMF from HI 21-cm galaxy surveys are restricted to $z \approx 0.1$ (e.g \citealt{hoppmann2015blind, ponomareva2023mightee}). In a recent work \cite{chowdhury2024hi} (referred as \citetalias{chowdhury2024hi} hereafter) presents an indirect estimate of the HIMF at $z \approx 1$, and finds that it shows a $\sim 4-5$ fold increase relative to that at $z \approx 0$ for galaxies with high HI masses. In this work, we provide an independent, indirect estimate of the $z \approx 1$ HIMF by jointly modeling   21-cm stacking observations of  the cosmological HI density parameter ($\Omega_\HI$; \citealt{chowdhury2020h}),  and the autocorrelation power spectrum (PS) of the HI 21-cm IM signal \citep{chime-auto-paper}.

Considering upgraded Giant Meterwave Radio Telescope  (uGMRT;
%\footnote{\url{https://www.gmrt.ncra.tifr.res.in/}}; 
\citealt{gupta2017upgraded}) observations, \cite{chowdhury2020h}  have stacked the 21-cm emission from $7653$ blue, star-forming galaxies at $z = 0.74 - 1.45$ in the DEEP2 field to estimate  $\Omega_\HI = (4.5 \pm 1.1) \times 10^{-4}$ at an average $z \approx 1.06$. There exist several other estimates of $\Omega_\HI$ from HI galaxy surveys at $z<0.2$ (see \citealt{padmanabhan2015theoretical} for a compilation) and DLA observations at $z \gtrsim 2$ (e.g \citealt{bird2017statistical, rhee2018neutral}).

Observations with the Canadian Hydrogen Intensity Mapping Experiment (CHIME;
% \footnote{\url{https://chime-experiment.ca/en}};
\citealt{bandura2014canadian}) have recently reported a measurement of the 21-cm autocorrelation PS at an average $z \approx 1.16$ \citep{chime-auto-paper}. The measured PS is shown by black points with $1 \sigma$ error bars in both panels of Fig.~\ref{fig:param_bins}. This measurement focuses on the frequency range $608.2-707.8 \, {\rm MHz}$, corresponding to the redshift interval $1.34 > z > 1.01$. The measurements were made in eight bandpowers spanning $0.4~h~{\rm Mpc}^{-1}\lesssim k \lesssim 1.5 ~h ~{\rm Mpc}^{-1}$, with a detection significance of $12.4 \, \sigma$. %The corresponding covariance matrices were estimated from $1000$ Gaussian noise realizations propagated through their full analysis pipeline.  
Several detections of the 21-cm PS have been reported at low redshifts using the MeerKAT
%\footnote{\url{https://www.sarao.ac.za/science/meerkat/about-meerkat/}} 
interferometer \citep{gibbon2015fiber}. \cite{paul2023first} claimed the first detection at $z = 0.32$ and $0.44$, followed by another recent detection by \cite{townsend2026measurements} at $0.02 \lesssim z \lesssim 0.07$.
The first attempts to detect the 21-cm PS at higher redshifts ($z = 1.32$) have been made using GMRT \citep{ghosh2011gmrt, ghosh2011improved}. A series of works \citep{chakraborty2021first, pal2022towards, elahi2023towards, elahi2023towards1, elahi2024towards} have reported an upper limit  $[\Omega_{\rm HI} b_{\rm HI}] < 0.011$ at $z = 2.28$ using uGMRT observations.

% Several detections of the 21-cm signal have also been made in cross-correlation with galaxy surveys. \cite{chang2010intensity} reported a detection of the IM signal at $z \sim 0.8$ using the data from the Green Bank Telescope (GBT\footnote{\url{https://public.nrao.edu/telescopes/gbt/}}) in cross-correlation with optical DEEP2 redshift survey. In a recent study by \cite{cunnington2023h}, observations from the MeerKAT interferometer in cross-correlation with WiggleZ Dark Energy \citep{drinkwater2010wigglez} Survey have been used to detect the signal at $0.4 < z < 0.45$. \cite{amiri2024detection} have recently reported a detection using the CHIME data in cross-correlation with Lyman-$\alpha$ forest from the Extended Baryon Oscillation Spectroscopic Survey (eBOSS; \citealt{du2020completed}).

The HI in the post-reionization era ($z \lesssim 5.5$) primarily resides in galaxies, which in-turn are hosted in dark matter halos. It is therefore appropriate to relate $M_\HI$ the HI mass contained in a galaxy to $M_h$ the mass of the halo through the HI mass - halo mass (HIHM) relation (e.g. \citealt{bagla2010h}). Here, we use the parametrization proposed by \cite{padmanabhan2017halo},
\begin{equation}
     M_\HI = \alpha f_{\rm H,c} \, M_h \left( \frac{M_h}{10^{11} h^{-1} M_\odot} \right)^\beta \exp \left[-\frac{M_{\rm cut}}{M_h}\right] \, ,
     \label{eq:HIHM}
\end{equation}
where
\begin{equation}
    M_{\rm cut} = 10^{10} M_\odot \left[\frac{v_{c0}}{60 \, {\rm km \, s^{-1}}} \right]^3   \left(\frac{1+z}{4} \right)^{-3/2} \, ,
    \label{eq:mcut}
\end{equation}
and $f_{\rm H,c} = 3 \Omega_b/ 4 \Omega_m$. This relation contains three free parameters, namely $\alpha$, $\beta$, and $v_{c0}$, corresponding respectively to the normalization, the excess logarithmic slope, and a lower cutoff on the circular velocity of the halo that can host HI. The HIHM relation provides a prescription to simulate the HI distribution by populating HI in the dark matter halos of a cosmological simulation.  

In this work, we constrain the parameters of the HIHM relation at $z \approx 1$ by fitting the simulated values of  $\Omega_\HI$ and the 21-cm PS to the measured values discussed earlier. We use the inferred HIHM relation to predict the $z \approx 1$ HIMF. Throughout our analysis, we have used the $\Lambda$CDM cosmological model with parameters taken from \cite{refId0}. 

%%%%%%%%%%%%%%%%%%%%%%%%%%%%%%%%%%%%%%%%%%%%%%%%%%%%%%%%%%%%%%%%%%%%%%%%%%%%%%%%%%%%%%%%%%%%%%%%%%%%%%%%%%%%%%%%%%%%%%%%%%%%%%%%%%%%%%%%%%

\section{Methodology}
\label{sec:method}

We have used a particle-mesh N-body code \citep{bharadwaj2004hi} to simulate the dark matter distribution at $z=1$, considering a comoving volume of $[150.08 \, \mpc]^3$ with $\approx 10^9$  particles of mass $1.089 \times 10^8 \, M_\odot$. The Friend-of-Friend algorithm was used to identify dark matter halos with a minimum halo mass $1.089 \times  10^9 M_\odot$, which corresponds to ten particles. The halos were populated with HI using the HIHM relation (eq.~\ref{eq:HIHM}), which were used to directly predict $\Omega_\HI$ and the  HIMF. The HI was assigned the peculiar velocity of the host halo (the HC method of \citealt{sarkar2018modelling}), and gridded to calculate the redshift space 21-cm PS $P(k)$  for the same $k$ bins as the observational data. We have used $50$ independent realizations of the simulations for the results presented here. These simulations are the same as those in \citet{chhabra2025probing}, which presents some more details of the simulation methodology. 
Note that while the HI simulation is performed at $z=1$, the measured $\Omega_\HI$ and 21-cm PS span a wider redshift range and are centered at $z = 1.06$ and $1.16$ respectively. We do not expect the HI distribution to evolve drastically in this small redshift interval. We assume that the systematic errors arising from the mismatch in the redshifts are negligible as compared to the observational uncertainties.

To constrain the HIHM parameters, we perform a joint Bayesian parameter estimation using the $z \approx 1$ observations of $\Omega_\HI$ and the  21-cm PS. We employ the affine-invariant Markov Chain Monte Carlo (MCMC) sampler implemented in the \texttt{emcee} package \citep{foreman2013emcee} to explore the 3D parameter space of our model. The likelihood is constructed assuming Gaussian-distributed measurement uncertainties and utilizes the covariance matrix provided with the CHIME PS data, which was estimated using 1000 noise-only simulations. In addition to this, the error covariance should also include the cosmic variance. While this can be theoretically predicted \citep{mondal2015effect}, here we have estimated this from our simulations and find it to be \re{several ($\sim 2-3$) orders of magnitude} smaller than the observational covariance provided with the observed 21-cm PS.

\begin{figure}
\centering
	\includegraphics[width=\columnwidth]{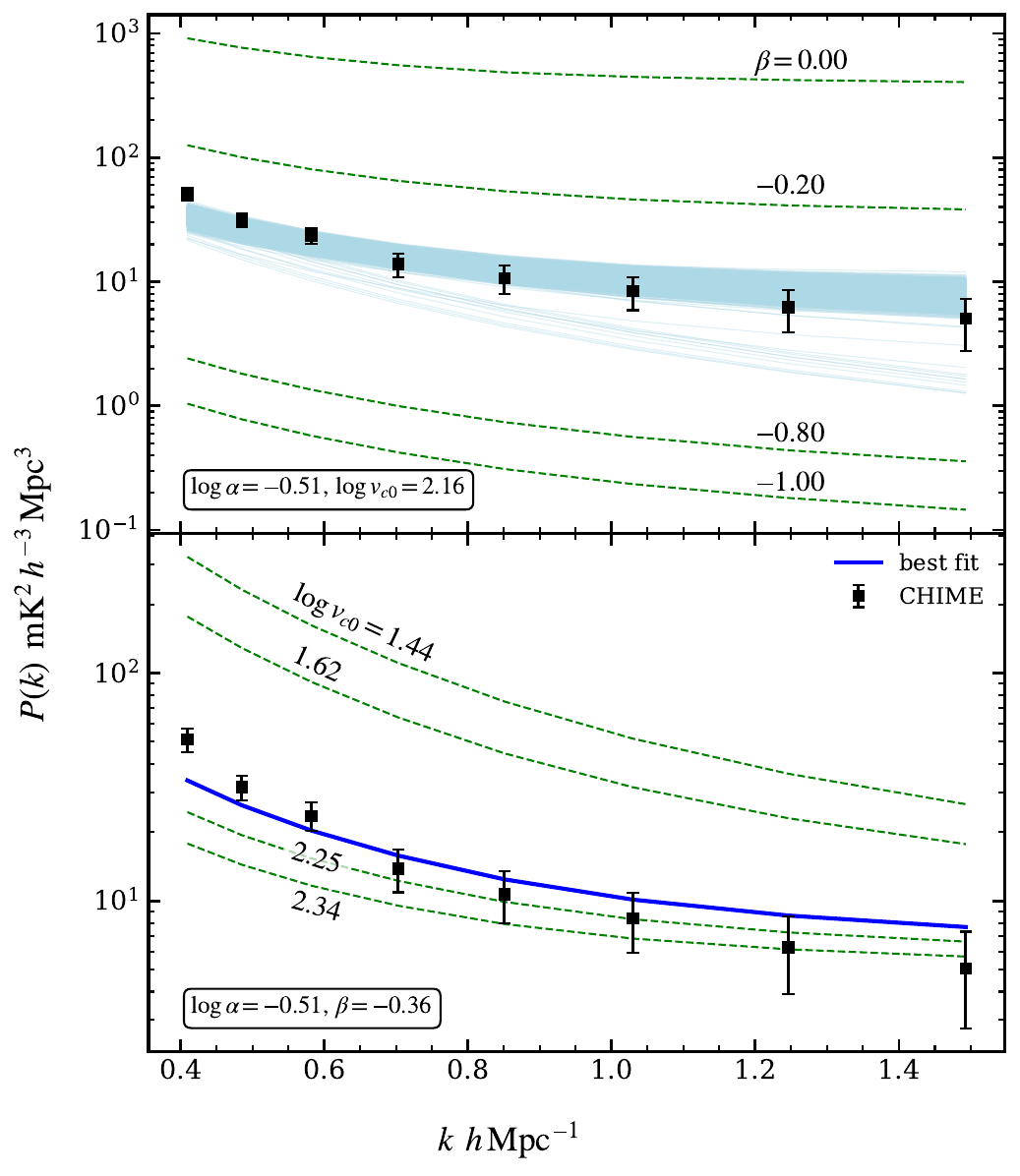} 
    \caption{The observed 21-cm PS by \citet{chime-auto-paper} is shown by black squares with $1 \sigma$ error bars. The dashed lines indicate the simulated PS as the HIHM parameters $\beta$ ({\it top}) and $v_{c0}$ ({\it bottom}) are changed, while the other parameters are kept fixed. The light blue curves in the top panel show the predicted 21-cm PS for 1000 randomly selected HIHM parameters from the posterior. The solid blue curve in the bottom panel shows the median of the sampled PS chosen to be the best-fit.}
    \label{fig:param_bins}
\end{figure}

The likelihood for a set of HIHM parameters $\theta \equiv (\log \alpha, \beta, \log v_{c0})$ consistent with the given measurements can be written as
\begin{equation}
    \label{eq:likelihood}
    \mathcal{L(\theta)} \propto \exp\left(-\frac{\chi^2}{2}\right),
\end{equation}
where
\begin{align}
    \begin{aligned}
        \chi^2 =  \sum_{i,j} & [P_{\rm obs}(k_i)-P(k_i, \theta)] \, C^{-1}_{ij} [P_{\rm obs}(k_j)-P(k_j, \theta)] \\
        &  + \, \left( \frac{[\Omega_{\rm HI}]_{\rm obs} - \Omega_{\rm HI} (\theta)}{\Delta \Omega_{\rm HI}} \right)^2 \, .
    \end{aligned}
    \label{eq:chi_sq}
\end{align}
Here,  $[\Omega_{\rm HI}]_{\rm obs}$  and $P_{\rm obs}(k)$  refer to the observed values whereas  $\Omega_{\rm HI} (\theta)$ and $P(k, \theta)$ refer to the predictions from the simulations, and $C_{ij}$ is the error covariance matrix for the observed 21-cm PS. 
% \re{Theoretically, the error covariance includes of the 21-cm PS includes contributions from higher-order terms like the Trispectrum \citep{mondal2015effect}. It is possible estimate these terms using higher-order perturbation theory, but this is beyond the scope of this work. However, to access their contribution in our analysis, we have estimated the covariance matrix using 50 independent realizations of our simulation for different values of the HIHM parameters. We find that the simulated covariance peaks at a value about $\sim 0.2 \, \times$ the observed covariance, and its incorporation in $\chi^2$ has a negligible effect on the results presented here.}
We have evaluated $\Omega_{\rm HI} (\theta)$ and  $P (k, \theta)$ from our simulations for the HIHM parameter values 
on an $11 \times 11$ grid spanning the range  $\beta \in [-1,0]$, and $\log v_{c0} \in [1.4,2.34]$ with $\alpha=\alpha_{\rm fix}=0.09$ fixed, and used linear interpolation to calculate $\Omega_{\rm HI} (\theta)$ and  $P (k, \theta)$ for other values of $\beta$ and $\log v_{c0}$.  Here, $\alpha$ is normalization factor, and we have scaled  $\Omega_{\rm HI} (\theta)$ and  $P (k, \theta)$ with $(\alpha/\alpha_{\rm fix})$ and $(\alpha/\alpha_{\rm fix})^2$ respectively to calculate the values for $\alpha$ in the range $\log \alpha \in [-2, 0]$. The dashed lines in Figure~\ref{fig:param_bins} shows  $P (k, \theta)$ for varying $\beta$ (top panel) and $v_{c0}$ (bottom panel). We see that the measured values $P_{\rm obs}(k)$ lie well within the range of $P (k, \theta)$ predicted by our simulations. Furthermore, we see that $P (k, \theta)$ show considerable variation as we vary the parameter values, indicating that it should be possible to constrain the parameter values using these observations. 

We adopt uniform priors over the parameter ranges indicated in Table~\ref{tab:bestfits} and verify that all chains have converged well before the final iteration. The best-fit values are chosen to be the medians of the 1D marginalized distributions for all parameters. 

%%%%%%%%%%%%%%%%%%%%%%%%%%%%%%%%%%%%%%%%%%%%%%%%%%%%%%%%%%%%%%%%%%%%%%%%%%%%%%%%%%%%%%%%%%%%%%%%%%%%%%%%%%%%%%%%%%%%%%%%%%%%%%%%%%%%%%%%%%

\section{Results}
\label{sec:results}

\begin{figure}
\centering
	\includegraphics[width=\columnwidth]{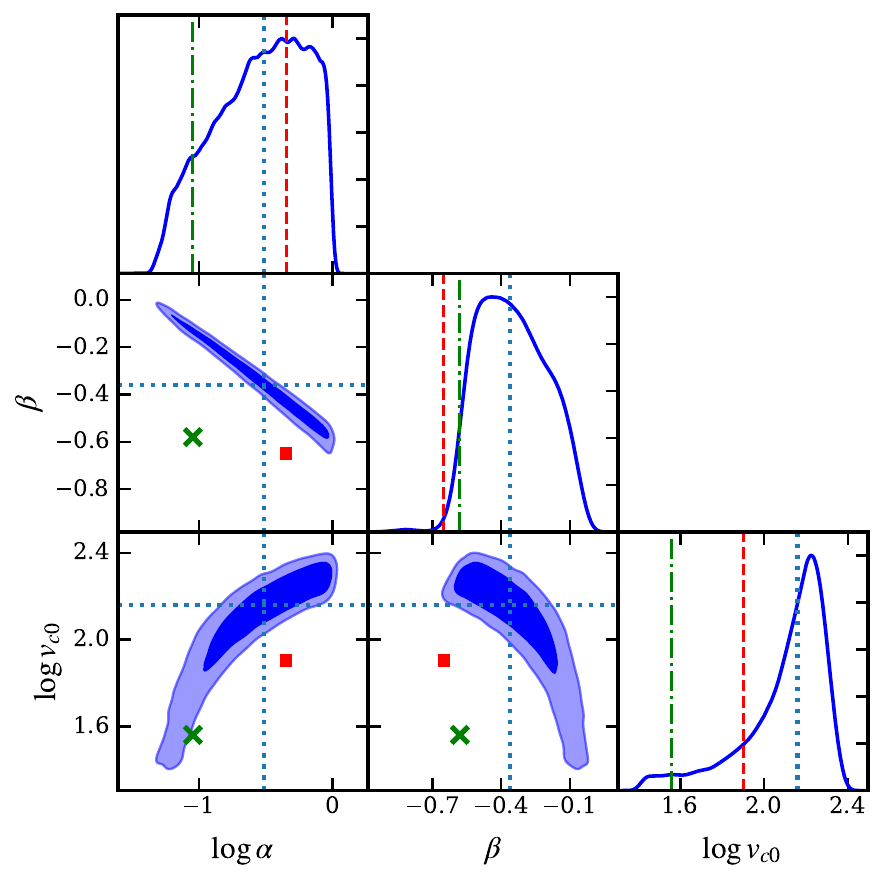} 
    \caption{Marginalized one and two-dimensional posterior distributions of the HIHM parameters,  
    using the $z \approx 1$ measurements of $\Omega_\HI$ and the 21-cm PS. The contours show the $68$th and $95$th percentiles. %confidence intervals of the 2D joint distributions and the histograms show the 1D marginalized distributions for the parameters. 
    The blue dotted lines indicate the best-fit values.% of the marginalized 1D distributions chosen to be the best-fit values. 
    The red square and dashed lines indicate the HIHM parameters fitted to obtain the HIMF by \citetalias{chowdhury2024hi}. The green cross and dot-dashed lines indicate the HIHM parameters estimated by \citet{padmanabhan2017halo}.}
    \label{fig:corner}
\end{figure}

Fig.~\ref{fig:corner} presents the marginalized one and two-dimensional posterior distributions of the three HIHM parameters.
The 2D contours show $68$ and $95$ percent credible intervals that are closed within the parameter ranges considered here. The shapes of the contours indicate strong correlations between the parameters. We see the parameters $(\log \alpha, \log v_{c0})$ are correlated, whereas $(\log \alpha, \beta)$ and $(\beta, \log v_{c0})$ are anti-correlated. Similar behavior is also observed when the likelihood analysis is performed using simulated PS and Bispectrum at large scales \citep{chhabra2025probing}. The 1D marginalized distributions are tightly constrained and have well-defined peaks. The blue dotted lines indicate the best-fit values of the HIHM parameters, which are also listed in Table~\ref{tab:bestfits} along with their $68$ percent credible intervals. We find here that best-fit values of the HIHM parameters are $\alpha = 0.30 \, ^{+ 0.36}_{- 0.19}$, $\beta = -0.36 \, ^{+0.18 }_{-0.15 }$ and $v_{c0} = 144 \, ^{+43}_{-62}$, which corresponds to a lower mass cut-off $M_{\rm cut} \approx 4 \times 10^{11} M_\odot$ on the halos that can host HI.
The light blue curves in the top panel of Fig.~\ref{fig:param_bins} show the predicted PS for $1000$ HIHM parameter values that are randomly drawn from the inferred posterior distribution. The blue solid curve in the bottom panel corresponds to the best-fit. We see that the observed PS (black points) lies well within the range of PS sampled from our posterior, except for the smallest $k$-bin, which lies $\sim 3\sigma$ away from our best-fit PS.

\begin{table}
    \label{tab:bestfits}
	\centering
	\caption{Prior ranges, best-fit values and $1 \sigma$ errors for the estimated HIHM parameters.}
	\label{tab:bestfits}
	\begin{tabular}{lcc} % four columns, alignment for each
		\hline
		Parameter & Prior & Best-fit \\
        \hline
        \\
        $\log \alpha$ & $[-2,0]$ &  $-0.51 \, ^{+ 0.33}_{-0.41}$ \\\\
		$\beta$ & $[-1,0]$ &  $-0.36 \, ^{+ 0.18}_{-0.15}$  \\\\
		$\log v_{c0}$ & $[1.4,2.34]$ &  $2.16 \, ^{+ 0.11}_{-0.24}$  \\\\
        \hline
	\end{tabular}
\end{table}

\begin{figure}
\centering
	\includegraphics[width=\columnwidth]{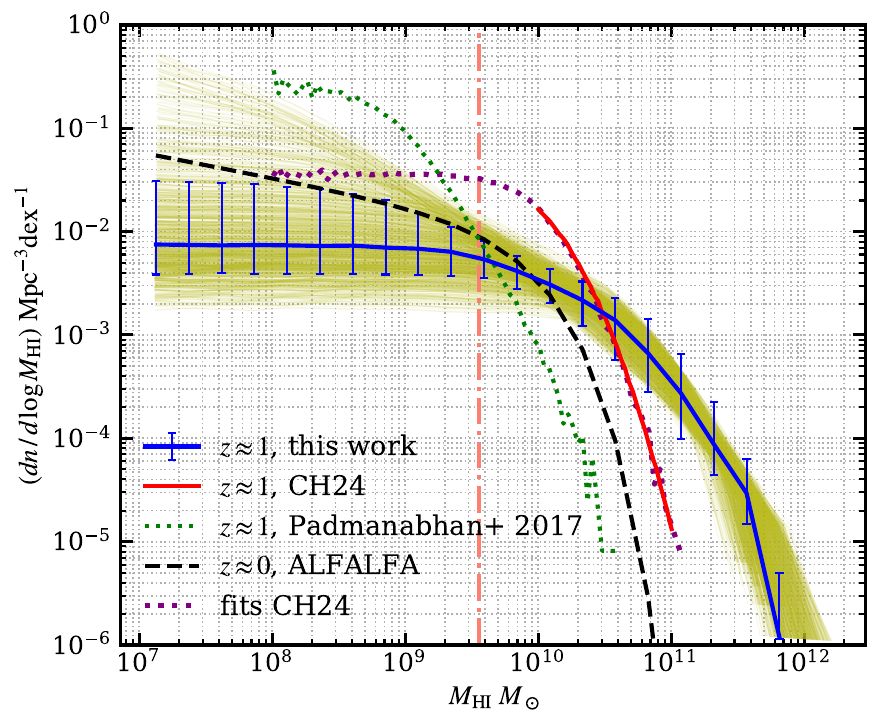} 
    \caption{$z \approx 1$ HIMF corresponding to $1000$ randomly selected HIHM parameters from the posterior (olive curves). The blue solid curve shows the median of the sampled HIMF, chosen to be the best-fit. The error bars indicate $68$ percent credible region around the best-fit. The orange dot-dashed line marks the HI mass corresponding to the best-fit lower mass cutoff $M_{\rm cut}$. The red solid curve shows $z \approx 1$ HIMF reported by \citetalias{chowdhury2024hi}. The $z \approx 1$ HIMF corresponding to HIHM parameters estimated by \citet{padmanabhan2017halo} is shown by the green dotted curve. The black dashed curve shows the $z \approx 0$ HIMF obtained using the ALFALFA survey \citep{jones2018alfalfa}.}
    \label{fig:h1mf}
\end{figure}

We now consider the HIMF for the HI distribution corresponding to the inferred posterior distribution. For a given set of HIHM parameters, we calculate the HIMF by counting the HI galaxies in bins of $\log M_\HI$. Fig.~\ref{fig:h1mf} shows the HIMF in the mass range $M_\HI = 10^7 - 10^{12} M_\odot$ estimated for a sample of $1000$ parameter values selected randomly from the posterior (olive curves). The best-fit HIMF (shown by blue solid curve) is chosen to be the median of the HIMF distribution corresponding the all the points in the posterior. The error bars indicate the $68$ percent credible region around the best-fit. We see that the HIMF remains constant at a value $\sim 7 \times 10^{-3} \, \mpc^{-3} \, {\rm dex}^{-1}$ for HI mass in the range $10^7 - 10^9 \, M_\odot$. However, this range makes an insignificant contribution to the total HI density. The HIMF declines gradually at larger HI masses till $7 \times  10^9 \, M_\odot$, beyond which it declines rapidly to $\sim 10^{-6} \, \mpc^{-3} \, {\rm dex}^{-1}$ as $M_\HI \rightarrow 10^{12} \, M_\odot$. We find that $\sim 90$ percent of the total HI gas is distributed primarily in the mass range $M_\HI \in [2 \times 10^9, \, 4 \times 10^{11}] \, M_\odot$. The red solid curve indicates the HIMF of star-forming galaxies at $z \approx 1$ by \citetalias{chowdhury2024hi}. We find that our HIMF predicts a larger abundance of  HI galaxies at the high mass end ($M_\HI \geq 3 \times 10^{10} \, M_\odot$). We also show the Schechter mass function corresponding to the $z \approx 0$ HIMF measured by the ALFALFA survey \citep{jones2018alfalfa}, indicated by the black dashed curve. 

%%%%%%%%%%%%%%%%%%%%%%%%%%%%%%%%%%%%%%%%%%%%%%%%%%%%%%%%%%%%%%%%%%%%%%%%%%%%%%%%%%%%%%%%%%%%%%%%%%%%%%%%%%%%%%%%%%%%%%%%%%%%%%%%%%%%%%%%%%

\section{Discussion}
\label{sec:summary}

The predicted HIMF (each olive curve in Figure \ref{fig:h1mf}) is the product of two quantities, namely the halo mass function (HMF; e.g. \citealt{sheth1999large}) which we determined directly from the cosmological N-body simulations and the HIHM relation whose parameters we have at our disposal. The blue solid curve in Fig.~\ref{fig:h1mf} shows the
predicted best-fit HIMF (median of the olive curves), which closely matches the HIMF for the best-fit parameter values shown in Table~\ref{tab:bestfits}. The purple dotted curve shows the HIMF for the parameter values $[\alpha, \beta, v_{c0}] = [0.45, \, -0.65, \, 80 \, {\rm km \, s}^{-1}]$, which provides a good fit to the HIMF measured by the \citetalias{chowdhury2024hi} (red solid curve). Both the blue and purple curves show a plateau region for low HI masses and a sharp drop at higher HI masses. The HI mass at which this drop starts is decided by the value of $v_{c0}$ or, equivalently, $M_{\rm cut}$, and this shifts to the left if $M_{\rm cut}$ is reduced. The orange dot-dashed vertical line indicates $M_\HI = 3.6 \times 10^9 \, M_\odot$, which corresponds to the halos of mass $M_{\rm cut} \approx 4 \times 10^{11} \, M_\odot$ for our best fit parameter values.  The plateau region corresponds to the HI contribution of the halos below the low-mass cutoff $M_{\rm cut}$, which makes an insignificant contribution to the total HI mass. Lowering the value of $v_{c0}$ increases the height of the plateau region. Lowering the value of $\beta$ causes the HIMF to drop more steeply for high HI masses. Increasing $\alpha$ causes the entire HIMF to shift to the right without affecting the shape. 

Compared to the \citetalias{chowdhury2024hi} HIMF, our estimate predicts a lower number density of low HI mass galaxies  ($M_\HI \lesssim 3 \times 10^{10} \, M_\odot$), and a higher number density of high HI mass galaxies. Our HIMF drops less steeply for high HI masses as compared to \citetalias{chowdhury2024hi}. Compared to hydrodynamical simulations such as TNG100, EAGLE, and SIMBA \citep{dave2020galaxy}, our HIMF predicts a higher number density of HI galaxies for $M_\HI \gtrsim 10^{10} \, M_\odot$ since the simulated HIMF drops even more steeply than the \citetalias{chowdhury2024hi} HIMF for high HI masses. It has been noted already that the measured 21-cm PS differs from the PS predicted by TNG100 simulation at $3.1 \sigma$ \citep{chime2026interpretation}. 

% However, recent 21-cm stacking observations using uGMRT \citep{chowdhury2022atomic} have shown that galaxies with $M_* \approx 10^{10} \, M_\odot$ become increasingly HI-dominated at $z \geq 1$, such that $\langle M_\HI \rangle / \langle M_* \rangle$ changes from $\sim 0.5, 1.4, 4.4$ at $z \approx 0, 1, 1.3$ respectively.

The red square and dashed lines in Fig.~\ref{fig:corner} indicate the HIHM parameter values that fit the \citetalias{chowdhury2024hi} HIMF measurement. In the 2D plots, we see that this point lies outside the $95$ percent credible contour of our inferred posterior distribution. From the 1D histograms, we note that these values are $[0.41 \sigma, \, 1.93 \sigma, \, 1.03 \sigma]$ away from the best-fit values of $[\alpha, \beta, v_{c0}]$ respectively. Also the corresponding cutoff halo mass is $6 \times 10^{10} \, M_\odot$, which is $0.17 \,$ times our best-fit $M_{\rm cut}$. In an earlier work, \citet{padmanabhan2017halo} have estimated the parameters of 
the HIHM  relation that we have considered here by combining the low-redshift observations of HI galaxy surveys ($z \approx 0$) and 21-cm intensity mapping ($z = 0.8$) with high redshift observations of DLA incidence and column density distributions ($z \sim 2.3 - 5$). They have used an analytical halo-model approach to obtain the parameters $\alpha = 0.09 \, \pm \, 0.01$, $\beta = -0.58 \, \pm \, 0.06$ and $\log v_{c0} = 1.56 \, \pm \, 0.04$, which corresponds to $v_{c0} = 36.3 \, {\rm km \, s}^{-1}$ and $M_{\rm cut} = 6.26 \times 10^9 \, M_\odot$ at $z=1$. The estimated values are indicated in Fig.~\ref{fig:corner} by the green cross and dot-dashed lines. The corresponding HIMF is shown by green dotted line Fig~\ref{fig:h1mf}, which predicts a larger number density of low HI mass galaxies  ($M_\HI \lesssim 3 \times 10^{9} \, M_\odot$), and a lower number density of high HI mass galaxies as compared to our HIMF and even that of \citetalias{chowdhury2024hi}. 

\re{
We note that the derived HIMF is potentially sensitive to the HIHM parametrization, which maps the underlying halo population to the HI distribution. Throughout the present analysis, we have adopted the parametrization by \cite{padmanabhan2017halo}, which consistently models the 21-cm PS and $\Omega_{\HI}$, and also allows us to infer the HIMF. Varying the functional form of the HIHM relation could modify both the normalization and shape of the inferred HIMF. In future work, we will investigate alternative HIHM parameterizations (e.g. \citealt{bagla2010h, padmanabhan2016constraining, villaescusa2018ingredients, baugh2019galaxy, spina2021h}) and quantify their impact on the inferred HIMF. We also note that the mass resolution of the halos in our simulations is another potential concern. However, we find that the preferred models have very little contribution from the low-mass halos since the best-fit $M_{\rm cut} = 4 \times 10^{11} \, M_\odot$ is much higher than the halo mass resolution ($\sim 10^9 \, M_\odot$). We quantify this resolution limitation by comparing the estimated $\Omega_\HI$ with an analytical prediction based on the halo model approach \citep{cooray2002halo}, and find that for the smallest $M_{\rm cut} = 2 \times 10^9 \, M_\odot$, the difference is about $\sim 17 \%$. However, the difference drops to $\sim 0.1 \%$ for the best-fit HIHM relation. We conclude that the low mass resolution has a negligible impact on our analysis.
}

\re{Our analysis predicts that $\sim 90$ percent of the total HI content resides primarily in the HI mass range $M_\HI \in [2 \times 10^9, \, 4 \times 10^{11}] \, M_\odot$, which can be mapped to the stellar mass range $M_* \in [5 \times 10^{9}, \, 5 \times 10^{11}] \, M_\odot$  using a widely accepted stellar mass - halo mass relation (SHMR; \citealt{girelli2020stellar}). \citet{lagos2014galaxies} have studied the contribution of galaxies with different properties to the global SFR, HI and molecular hydrogen ($H_2$) densities as a function of redshift using the \texttt{GALFORM} semi-analytical galaxy formation model \citep{cole2000hierarchical}. They conclude that the $H_2$ density is dominated by galaxies with high SFR $> 1 \, M_\odot \, {\rm yr}^{-1}$ and high stellar masses $M_* > 10^{10} \, M_\odot$, while the HI density is always dominated by galaxies with low SFR $< 1 \, M_\odot \, {\rm yr}^{-1}$ and low stellar masses $M_* < 10^9 \, M_\odot$, which is in contrast with our prediction. A similar discrepancy has been noted in previous 21-cm stacking studies at $z < 0.11$ \citep{hu2020cosmic}, which show that galaxies with $M_*$ above and below $\sim 2 \times 10^9 \, M_\odot$ contribute equally to the total HI content. \cite{hu2020cosmic} claim that this discrepancy may be partly due to the HI/$H_2$ partitioning recipes utilized in the semi-analytical galaxy formation models, and there may be a wider overlap between galaxy populations dominating the HI and $H_2$ densities. Our results indicate that this tension persists and sharpens at $z \approx 1$. Direct HI 21-cm measurements of individual, stellar-mass-selected galaxies at $z \approx 1$ will be needed to resolve the inconsistency.}

% \rg{We also note that the derived HIMF is potentially sensitive to the HIHM parametrization, which maps the underlying halo population to the HI distribution. Throughout the present analysis, we have adopted the parametrization by \cite{padmanabhan2017halo}, which consistently models the 21-cm PS and $\Omega_{\HI}$, and also allows us to infer the HIMF. Varying the functional form of the HIHM relation could modify both the normalization and shape of the inferred HIMF. In future work, we will investigate alternative HIHM parameterizations (e.g. \citealt{bagla2010h, padmanabhan2016constraining, villaescusa2018ingredients, baugh2019galaxy, spina2021h}) and quantify their impact on the inferred HIMF. }

Our analysis predicts a larger abundance of high mass HI galaxies at $z \approx 1$ as compared to earlier estimates. These predictions have significant implications for the study of galaxy evolution and our understanding of how star formation occurs. Future observations with CHIME, MeerKAT and the upcoming SKA-Mid\footnote{\url{https://www.skao.int/en/explore/telescopes/ska-mid}} are expected to yield more precise estimates of $\Omega_\HI$, 21-cm PS and also the 21-cm Bispectrum. It is anticipated that such observations would allow us to probe the high redshift HI distribution in more detail \citep{chhabra2025probing}. The formalism described in this work can be applied for any tracer of the underlying matter field whose distribution is dependent on the masses of their host dark matter halos. Synergies of tracers such as CO/[CII]/[OIII] lines (see \citealt{bernal2022line}) that directly trace the Interstellar medium (ISM), with the 21-cm signal may provide a more complete picture of cosmic star-formation at the post-EoR redshifts.
An understanding of the high redshift HI distribution will also help towards using 21-cm IM to study large-scale structure formation and cosmological parameter estimation.

%%%%%%%%%%%%%%%%%%%%%%%%%%%%%%%%%%%%%%%%%%%%%%%%%%%%%%%%%%%%%%%%%%%%%%%%%%%%%%%%%%%%%%%%%%%%%%%%%%%%%%%%%%%%%%%%%%%%%%%%

\section*{Acknowledgments}
We acknowledge the computing facilities in the Department of Physics, IIT Kharagpur. MC acknowledges the support of the Prime Minister Research Fellowship (PMRF).

\section*{Data Availability}
The simulated data and packages used in this work will be shared upon reasonable request to the authors.

\bibliography{refer_old}{}
\bibliographystyle{aasjournal}
\label{lastpage}

\end{document}